\documentclass{acm_proc_article-sp}
\usepackage{graphicx,subcaption,booktabs,hyperref}
\usepackage{epstopdf}
\graphicspath{{D:/Research/DSSG/Housing_Justice_V2/ADT/figs/}}
\usepackage[table]{xcolor}
\usepackage{color}
\hypersetup{colorlinks=true,linkbordercolor={1 1 1},citecolor=black,linkcolor=black,urlcolor=black}

\begin{document}

\conferenceinfo{}{Bloomberg Data for Good Exchange 2016, NY, USA}

\title{Using data science as a community advocacy tool to promote equity in urban renewal programs: An analysis of Atlanta's Anti-Displacement Tax Fund}

\numberofauthors{9}
\author{
\alignauthor
Jeremy Auerbach\\
       \affaddr{University of Tennessee}\\
       \affaddr{Knoxville, TN}\\
       \email{jauerbac@vols.utk.edu}
\alignauthor
Hayley Barton\\
       \affaddr{Duke University}\\
       \affaddr{Durham, NC}\\
       \email{hayley.barton@duke.edu}
\alignauthor
Takeria Blunt\\
       \affaddr{Spelman College}\\
       \affaddr{Atlanta, GA}\\
       \email{blunttakeria@gmail.com}
\and
\alignauthor
Vishwamitra Chaganti\\
       \affaddr{Georgia State University}\\
       \affaddr{Atlanta, GA}\\
       \email{vchaganti2@student.gsu.edu}
\alignauthor
Bhavya Ghai\\
       \affaddr{Stony Brook University}\\
       \affaddr{Stony Brook, NY}\\
       \email{bghai@cs.stonybrook.edu}
\alignauthor
Amanda Meng\\
       \affaddr{Georgia Institute of Technology}\\
       \affaddr{Atlanta, GA}\\
       \email{a.meng@gatech.edu}
\and
\alignauthor
Christopher Blackburn\\
       \affaddr{Georgia Institute of Technology}\\
       \affaddr{Atlanta, GA}\\
       \email{cblackburn8@gatech.edu}
\alignauthor
Ellen Zegura\\
       \affaddr{Georgia Institute of Technology}\\
       \affaddr{Atlanta, GA}\\
       \email{ewz@cc.gatech.edu}
\alignauthor
Pamela Flores\\
       \affaddr{HELP Organization Incorporate}\\
       \affaddr{Atlanta, GA}\\
       \email{pamela@helporginc.org}
}

\maketitle
\begin{abstract}
Cities across the United States are undergoing great transformation and urban growth. Data and data analysis has become an essential element of urban planning as cities use data to plan land use and development. One great challenge is to use the tools of data science to promote equity along with growth. The city of Atlanta is an example site of large-scale urban renewal that aims to engage in development without displacement. On the Westside of downtown Atlanta, the construction of the new Mercedes-Benz Stadium and the conversion of an underutilized rail-line into a multi-use trail may result in increased property values. In response to community residents' concerns and a commitment to development without displacement, the city and philanthropic partners announced an Anti-Displacement Tax Fund to subsidize future property tax increases of owner occupants for the next twenty years. To achieve greater transparency, accountability, and impact, residents expressed a desire for a tool that would help them determine eligibility and quantify this commitment. In support of this goal, we use machine learning techniques to analyze historical tax assessment and predict future tax assessments. We then apply eligibility estimates to our predictions to estimate the total cost for the first seven years of the program. These forecasts are also incorporated into an interactive tool for community residents to determine their eligibility for the fund and the expected increase in their home value over the next seven years.
\end{abstract}





\section{Introduction}

Municipalities and companies that work with big data are making incredible strides toward using data effectively; however, the people that the data represent can get lost in the numbers, diminishing their importance. Bridging the gap between community members and entities that use big data for decision making is a crucial and ever-evolving process. In Atlanta, the ongoing construction of the new Mercedes-Benz Stadium and Beltline trail has left Atlanta's Westside residents feeling excluded and frustrated, with fears that the new development that taxpayers shared the cost to build will expedite displacement and accelerate gentrification\footnote{In a conversation with Westside residents.}.

As the back-to-the-city movement continues across the United States, cities have undertaken major urban improvement and revitalization projects that have resulted in increased rents and property appreciation. For example, home value appreciation due to new rail networks has been observed in Chicago \cite{mcm04}, Boston \cite{arms94} and Portland \cite{chen98}. Atlanta is currently undergoing a city-wide transportation revitalization project through the conversion of an underutilized rail-line into a 22-mile multi-use bike and pedestrian trail, called the Beltline \cite{atlblt17}. Since construction on part of Atlanta's Eastside Beltline has been completed, increased rents and property appreciation have been observed \cite{imm09}. As construction continues on the Westside Beltline segments, nearby residents are concerned about possible increase in rents and property taxes.

In addition to development spurred by the Beltline, Atlanta's Westside neighborhoods are adjacent to the newly constructed Mercedes-Benz Stadium, a multi-billion dollar sports and entertainment venue. This development has been quite contentious from the start. Residents question the use of public funds for a sports and entertainment venue and have already experienced displacement, as the development is located adjacent to the Georgia Dome which is on the footprint of a now extinguished community, Lighting Alley, and most recently required the demolition of two historically black churches \cite{bels17}. In response to these concerns, the city, along with the private sector, created an Anti-Displacement Tax Fund. This fund is a collaboration between the city and the Westside Future Fund, a non-profit organization, which is administering the fund and has contracted a local consulting group to aid in executing the fund \cite{wsff17b}. The Anti-Displacement Tax Fund will attempt to offset the increase in property taxes for eligible homeowners in four neighborhoods on Atlanta's Westside as home values rise (see Table \ref{elig} for the eligibility requirements).

Other cities have tried to predict displacement and gentrification, including Los Angeles and San Francisco \cite{lait16,ud17}. Still, there has been very little work examining the cost and impact of suggested anti-displacement programs. In Atlanta, the only public estimate of the cost and scope of the Tax Fund predicted that 400 homeowners in the Westside neighborhoods would participate, and 165 would remain in their homes at the end of twenty years for a total program cost of five million dollars over twenty years \cite{bed17}. We reevaluated this study by including machine learning techniques, a richer dataset, and household-level eligibility.

We partnered with a local community group that seeks to improve the quality of life of in-place residents by mobilizing a community-driven land trust program (\href{http://waltprogram.org}{waltprogram.org}). This partnership between Georgia Tech and WALT has been established with the goal of using data to estimate and share with community members the cost and impact of a new Anti-Displacement Tax Fund, which is described in the following section. Our process, documented in Section 3., and the tool that we created (Section 4.) demonstrate how community engagement and data science can be combined to examine programs and policies, and most importantly, how community members can be informed and involved in this process to increase transparency and accountability in public policy (discussed in Section 5.). Finally, we discuss the implications of our findings and future work in Section 6.

\section{Current Approach}

The original estimates for the program eligibility and cost of the Anti-Displacement Tax Fund used aggregated Census data and several assumptions when data was not available \cite{bed17}. Using aggregate data and assumptions could impact the accuracy of the projections. To estimate the number of owner-occupants in the area the model used the rate of residents claiming a homestead tax exemption, which requires the homeowner to also be the occupant of the residence. The homestead exemption rate for the Westside neighborhoods was 79\% in 2016. This rate was also used as the program participation rate and the model assumed a 5\% annual dropout rate. As we show in the following section, we find this exemption rate significantly overestimates the actual owner-occupancy rate.

\begin{table}[ht!]
\centering
\begin{tabular}{|l|p{3.8cm}|}
\hline
Location             & Participants must reside in one of the following Westside neighborhoods: English Avenue, Vine City, Atlanta University Center, and Ashview Heights.$^{*}$                                                     \\ \hline
Enrollment Period	 & Residents may apply for the program's first year between April 12, 2017 and March 15, 2018. Participants will need to reapply annually between January 1 and March 15 to continue receiving program funding.
\\ \hline
Income               & Participants' household income must be below the Area Median Income (AMI) for their household size.                                                                                                      \\ \hline
Owner-Occupancy      & Only homeowners who both currently live in their home and have lived there for at least one year are eligible. New homeowners are also eligible if the home had been previously enrolled in the program. \\ \hline
Liens/Property Taxes & Participants cannot have any standing liens or debts attached to their property.                                                                                                            \\ \hline
Heirs                & Heirs who meet all other conditions are also eligible.                                                                                                                                                   \\ \hline
\end{tabular}
\caption{Summary of Anti-Displacement Tax Fund Eligibility Requirements. *We have also included the Washington Park neighborhood in pieces of our analysis at the request of community members, to determine the cost and feasibility of adding this area to the program boundaries.}
\label{elig}
\end{table}

Another major eligibility requirement, income, was estimated from aggregated Census data with 62\% of the households being eligible. However, the program affords for income levels at different household sizes. The original model failed to account for household size, which affects the income thresholds. We find that more households would be eligible when this is accounted for. We also found issues with the property value appreciation model. For property value appreciation, the original model assumed properties valued at \$37,000 would appreciate annually at 12\%. Properties valued under the \$37,000 were assumed to appreciate at 50\% until it reached the \$37,000 threshold, after which it was set to an annual rate of 12\%. This 12\% was taken from the average property appreciation rates found in other Atlanta neighborhoods since 2012. This is a broad assumption as these other Atlanta neighborhoods are not socioeconomically similar to the Westside neighborhoods in the affected area and have not experienced significant development projects. Additionally, property appreciation for Atlanta neighborhoods that have undergone a recent development project have not followed these simple trends (as we show in the next section). The original study estimated 400 residents would be eligible and enroll with a program cost of \$0.5M for the first 7 years and \$5M over 20 years. Our approach finds this modeling considerably underestimated the number of eligible residents and program costs.

\section{Our Approach}

We take a two-fold approach to quantify the expected cost of the Anti-Displacement Tax Fund. First, we developed a method that determined the number of homeowners who would be eligible. Second, we developed a method for predicting the assessed values of the properties in each of the qualifying neighborhoods. One of the key innovations of our study is the use of publicly available data to determine household level income and projected property appreciation, methods that are easily accessible for analysts to replicate. Consequently, our study provides a core framework that can be replicated for analysis of property tax subsidization policy programs in other locales across the country.

\subsection{Data}

We relied on publicly available data from various local and national government agencies and Zillow.com to conduct our program analysis. First, we used the historical Fulton County Tax Assessor data for individual home characteristics and historical tax assessments from 2005-2016 \cite{fcba17}. Second, we used data collected from the Georgia Superior Court Clerks' Cooperative Authority database to determine homeowner liens in the program area \cite{gsccca17}. Third, our income prediction model utilized data from the U.S. Bureau of Labor Statistics Consumer Expenditure Survey for the years 2013, 2014, and 2015, scraped 2017 Zillow rent estimate data, and 2015 American Community Survey neighborhood population estimates \cite{bls16,zill17,uscb15}.\footnote{The Zillow rent estimate data was collected in July 2017 and represents the company's best estimates for monthly rental payments at the household level.} The CEX was a random survey of residents across the U.S. and the dataset includes the following household-level attributes that were used in the modeling: before-tax-income, monthly rent payments, the number of bedrooms, the number of bathrooms, the number of rooms, and the age of the house. We used the low rent estimates from Zillow to get a more liberal estimate of the total cost of the program, as more households would qualify for the program.

Despite the richness of the data used in our study, we encountered several issues while working with this data, finding numerous discrepancies and missing values within the Fulton County Tax Assessor data, including the lack of 2009 data. This is not an unexpected hurdle as there are over 100,000 residential parcels in Fulton County and approximately 20 tax assessors. Furthermore, the lien data was stored in formats that were not machine readable and required a considerable amount of time to parse. Even with these issues, we were able to collect a substantive amount of usable data in a short time period for the analysis.

\subsection{Program Eligibility Estimation}

Determining the individual households eligible for the Anti-Displacement Tax Fund was crucial both to estimate the total cost of the program and for our interactive eligibility tool. As mentioned in Table ~\ref{elig}, we identified the eligible households based on location, owner occupancy, the presence of liens, and income (which is dependent on the household size).

{\bf Location:} Using the Fulton County Tax Assessor 2017 data and we filtered out all non-empty lot residential parcels which lie in Ashview Heights, Atlanta University Center, Vine City, English Avenue or Washington Park neighborhoods, which resulted in ~2600 geographically eligible residential parcels.

{\bf Owner occupancy:} From the Fulton County Tax Assessor 2017 data, we compared parcel addresses with owner addresses to determine if the household is owner occupied. We also considered all households who claim Homestead exemption, which requires the owner to be the occupant, and ~36\% of the homes were found to be owner occupant.

{\bf Lien status:} Lien data were gathered from the Georgia Superior Court Clerks' Cooperative Authority. Since the data was not machine readable, a random sample of ~30\% homeowners in the four original Westside neighborhoods and ~45\% Washington Park were gathered. Of the households in the four Westside neighborhoods, ~59\% did not have liens, nor did ~58\% in Washington Park.

{\bf Income:} Based on ZIP code-level tax return data from 2014, we found that ~90\% of households in the Westside might qualify for the program based on household income which is significantly less than \$47,250 \cite{irs14}. A simple assumption might be to consider all households eligible based on income criterion, but we attempted to predict household-level income eligibility based on observable, physical characteristics of a house and its expected rent. Since individual household income data was unavailable, we used the relationship for rent estimates and household income. For each house in the program area, we merged the house characteristics from the Fulton County Tax Assessor data and the Zillow rent estimates for each residential property in the Westside.

To model the relationship between rent and income, we first filtered the Consumer Expenditure Survey (CEX) data by the Atlanta homeowner participants, resulting in ~250 survey participants, and observed a strong correlation between rent and income. We also observed a large number of missing values in the data, specifically low and high incomes were statistically absent. This is a known phenomena for income surveys \cite{groves98}. Prior to modeling the relationship between rent and income, we used random forests to impute missing values. Random forests were able to adequately deal with the collinearity between variables, overfitting from the large number of variables (over 300), the presence of nonlinear relationships, heteroskedasticity, and non-normal distributions of the data. Once the imputation was satisfactory and assuming that the patterns learned from the Atlanta CEX dataset closely resemble the Westside, we used random forests to predict household incomes for each of the Westside homes.

To estimate the household size, we used interpolated population estimates based on the 2015 American Community Survey for the Atlanta neighborhoods in the program area. We used the Fulton County Tax Assessor data to find the average number of bedrooms for each of the eligible neighborhoods. Finally, we divided the population estimates with the number of rooms in each neighborhood to find the average number of residents per bedroom. Table 2 provides a summary of these estimates for each of the neighborhoods in the program area, including Washington Park, and ~92\% of the Westside households satisfied the income eligibility.

\begin{figure}[b!]
\centering
\begin{subfigure}{0.475\textwidth}
  \centering
  \includegraphics[trim = 0mm 0mm 0mm 0mm, clip,width=\linewidth]{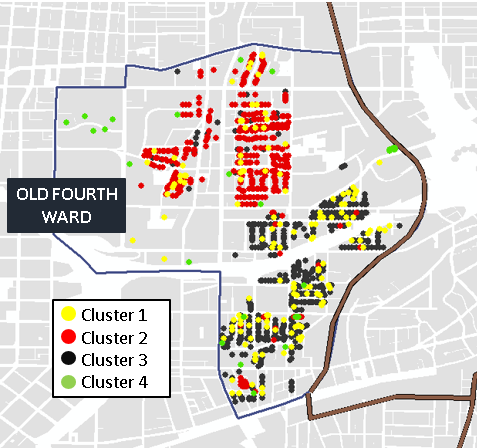}
\end{subfigure}
\\
\vspace{1mm}
\begin{subfigure}{0.475\textwidth}
  \centering
  \includegraphics[trim = 0mm 0mm 0mm 0mm, clip,width=\linewidth]{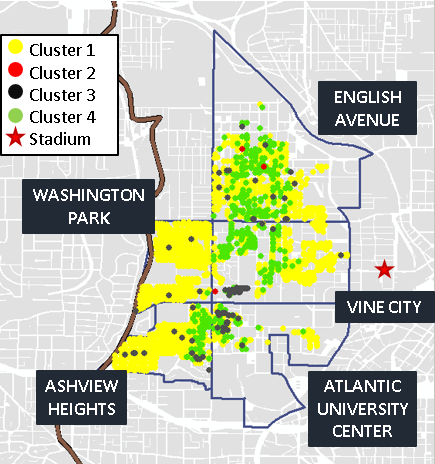}
\end{subfigure}
\caption{The top map displays the residential parcels in the Old Fourth Ward (Eastside Atlanta) clustered on the historical tax assessment time series (2005-2016). The bottom map shows the residential parcels for the the five Westside neighborhoods nearby the current Beltline construction similarly clustered. The Beltline is denoted by the brown line and neighborhoods are bordered blue.}
\label{maps}
\end{figure}

\subsection{Tax Assessment Forecasting}

Although the new Mercedes-Benz Stadium may impact Westside Atlanta home values, previous studies of sports venues have found mixed effects on property values. A study of Texas metropolitan areas found that property values decreased after new sports venue announcements \cite{dehring07}. Another study found positive price improvement near the construction of the FedEx stadium in Landover, MD \cite{tu05}. Because of these conflicting findings, we excluded potential stadium-related effects in our estimates of home values and, instead, focused on the effect of the Beltline trail construction, which has been shown to increase the property values in neighborhoods around proposed and completed segments \cite{imm17}.

To estimate future tax assessments of homes in the Westside, we began with the assumption that the Westside would experience changes similar to the impact of completed BeltLine construction near the Old Fourth Ward neighborhood. This assumption is validated primarily by shared characteristics, including proximity to urban renewal projects, proximity to Atlanta's downtown, proximity to industrial land use zones, and similar historic socioeconomic makeup. For this analysis, we gathered historical tax assessment data from 2005 through 2016 for each parcel from the Fulton County Tax Assessor's Office, with the exception of data from 2009 which were unavailable at the time of the analysis. Once the data were cleaned, we found 2415 complete residential time series, which were converted to percent-change differences to ensure property appreciation trends for homes with different nominal property values could be compared.

\begin{figure}[h]
\centering
  \includegraphics[trim = 0mm 5mm 5mm 20mm, clip,width=0.475\textwidth]{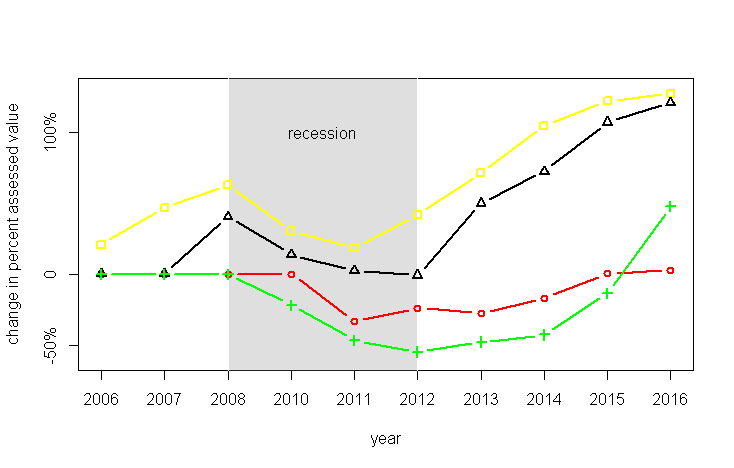} 
\caption{Average cumulative change in percent tax assessment for the clusters of the historical tax assessment time series in the Old Fourth Ward neighborhood (Eastside Atlanta) for the period 2005-2016. The data were first converted to percent differences and 2009 data were not available.}
\label{clusters}
\end{figure}

Due to the spatial and temporal variation and nonlinearity in the tax assessment time series tends, using a single static trend statistic, as the 2017 study did, was considered inappropriate. We clustered the time series with a binary distance matrix using Ward's method, to ensure variation in each cluster was minimized while variation between each was maximized \cite{ward63}. We chose a binary distance measure to detect even minor differences in time series variation, as the tax assessments for a specific parcel are generally conducted every 3 to 4 years and regularly showed small changes. We observed four significant clusters, with 876, 224, 251, and 1064 homes in each cluster (see Figure  \ref{maps} and Figure \ref{clusters}). Once we created these time series clusters, we used random forests to identify the most influential housing characteristics \cite{brei01}. The most important features that determine property appreciation trends were the distance to the Beltline, property value in 2005 (the year before construction of the Beltline) and age of the home (see Table \ref{features}) . We chose random forests to identify the impact of home features on the clusters to account for the nonlinear relationships present between the tax assessments and home features, and to avoid overfitting the data due to strong correlations between home characteristics.

\begin{table}[h!]
\centering
\begin{tabular}{l p{0.8cm}p{0.8cm}p{0.8cm}p{0.8cm}}
\toprule
Feature                & Cluster 1               & Cluster 2               & Cluster 3               & Cluster 4               \\ \midrule
Property value in 2005 & \cellcolor{darkgray}  & \cellcolor{black}     & \cellcolor{black}     & \cellcolor{darkgray}  \\
Land use - 1 family    & \cellcolor{gray}      & \cellcolor{gray}      &                         & \cellcolor{lightgray} \\
Land use - 2 family    & \cellcolor{lightgray} &                         &                         &                         \\
Land use - 3 family    &                         &                         &                         &                         \\
Land use - condo       & \cellcolor{gray}      & \cellcolor{darkgray}  & \cellcolor{lightgray} & \cellcolor{lightgray} \\
Land use - townhouse   & \cellcolor{gray}      & \cellcolor{gray}      & \cellcolor{gray}      & \cellcolor{lightgray} \\
Land use - condo loft  & \cellcolor{gray}      & \cellcolor{lightgray} & \cellcolor{lightgray} & \cellcolor{lightgray} \\
Living units = 1       & \cellcolor{lightgray} &                         &                         &                         \\
Living units = 2       & \cellcolor{lightgray} &                         &                         &                         \\
Living units = 3       &                         &                         &                         &                         \\
Land size (acres)      & \cellcolor{darkgray}  & \cellcolor{darkgray}  & \cellcolor{gray}      & \cellcolor{gray}      \\
Home size (sq ft)      & \cellcolor{darkgray}  & \cellcolor{darkgray}  & \cellcolor{darkgray}  & \cellcolor{gray}      \\
Number of rooms        & \cellcolor{gray}      & \cellcolor{darkgray}  & \cellcolor{gray}      & \cellcolor{lightgray} \\
Number of bedrooms     & \cellcolor{gray}      & \cellcolor{lightgray} &                         & \cellcolor{lightgray} \\
Number of bathrooms    & \cellcolor{gray}      & \cellcolor{lightgray} & \cellcolor{lightgray} & \cellcolor{lightgray} \\
Distance to BeltLine   & \cellcolor{black}     & \cellcolor{black}     & \cellcolor{darkgray}  & \cellcolor{black}     \\
Atlanta exemption      & \cellcolor{lightgray} & \cellcolor{gray}      &                         &                         \\
Fulton exemption       & \cellcolor{lightgray} & \cellcolor{gray}      &                         &                         \\
Owner occupied         & \cellcolor{lightgray} & \cellcolor{gray}      &                         &                         \\
Age of home            & \cellcolor{darkgray}  & \cellcolor{black}     & \cellcolor{gray}      & \cellcolor{gray}      \\ \bottomrule
\end{tabular}
\caption{Important housing characteristics for each cluster in Old Fourth Ward (Eastside Atlanta). Darker shades represent more important features. Distance to the Beltline, the property value in 2005, and the age of the home are the primary features for variation in the historical tax assessment time series trends.}
\label{features}
\end{table}

Westside residential parcel characteristics were also sourced from the Fulton County Tax Assessor data for the following five Atlanta neighborhoods: Ashview Heights, Atlanta University Center, English Avenue, Vine City, and Washington Park. We fit the 2562 identified parcels into the clusters generated from the Old Fourth Ward neighborhood though random forest testing (see Figure \ref{maps}). To forecast the future tax assessment values for homes in the Westside, we used the corresponding time series trends from the Old Fourth Ward neighborhood clusters after removing the recession time period data (2008-2012), which resulted in projections to 2024.

\subsection{Program Cost Estimation}

We conducted program costs estimates for several different scenarios: (i) accounting for lien rates and with and without residential parcels in Washington Park, (ii) disregarding liens rates with and without Washington Park, and (iii) including a 5\% dropout rate and a 79\% enrollment rate for all of the previous scenarios (which the original study incorporated and we considered for comparison). The Fulton County and City of Atlanta 2016 Millage rates were applied to tax assessment estimates after individual household exemptions were identified. Final program costs were calculated by summing all of the future property taxes minus the 2017 taxes for eligible households over the 7-year period. To simulate liens for households which we did not have data, we used Monte Carlo methods with the sample lien rates for the neighborhoods. The distribution of program cost estimates was normal so the average program cost estimate was used in the final results. Monte Carlo methods were likewise used for estimating the program costs with when household dropout and enrollment was considered.

\subsection{Online Tool}

\begin{figure}[h]
\centering
  \includegraphics[trim = 0mm 0mm 0mm 0mm, clip,width=0.475\textwidth]{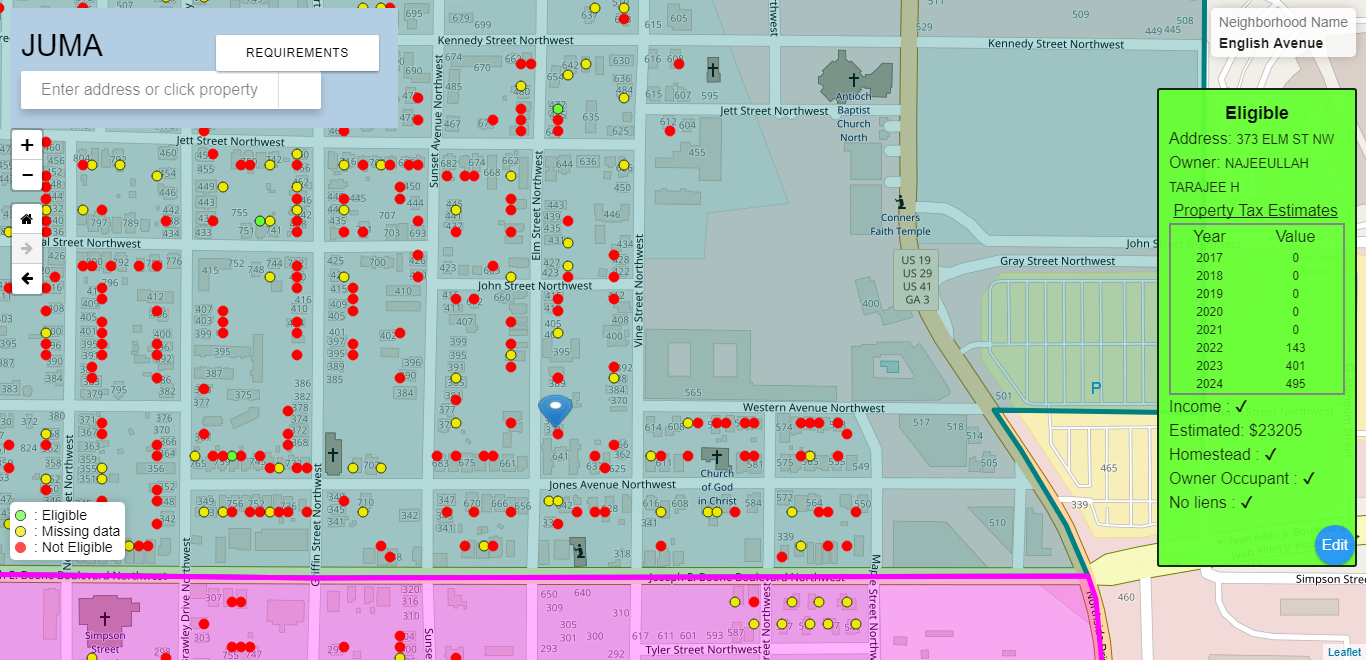} 
\caption{Web Mapping tool used to inform residents on their eligibility status and projected home value appraisals to the year 2024.}
\label{map}
\end{figure}

One of the primary goals of this research is to ensure that Westside community members are aware of the eligibility requirements of the fund. To assist residents in making an informed decision on participating in the fund, we developed an interactive web tool for Westside homeowners to view eligibility information and projected property taxes over the next seven years. (Figure \ref{map} and \href{http://dssg.gatech.edu/adt/}{dssg.gatech.edu/adt/}). Our partners at the Westside Atlanta Land Trust will assist us in introducing the tool and the information it provides to residents in order to spread awareness about the program's impact. Through this canvassing, they can also update eligibility information within the tool to provide better program estimates and validate the modeling, but issues with residents sharing personal information and validating any responses are still expected.

\section{Preliminary Results}

\begin{table*}[ht!]
\centering
\begin{tabular}{lcccc}
\toprule
& \multicolumn{2}{c}{Eligible Households} & \multicolumn{2}{c}{Total Program Cost} \\ \cmidrule{2-5}
& With Liens & Without Liens & With Liens & Without Liens                          \\
\midrule
Without Dropout \& Full Enrollment & & & &\\
\midrule
With Washington Park      & 489    & 702    & 2.7M    & 4.0M                              \\
Without Washington Park    & 372  & 560   & 2.1M    & 3.2M                              \\
\midrule
With 5\% Dropout \& Full Enrollment & & & &\\
\midrule                           
With Washington Park & 489$\rightarrow$339  & 702$\rightarrow$487    & 2.0M   & 3.1M                              \\
Without Washington Park   & 372$\rightarrow$257  & 560$\rightarrow$389 & 1.6M  & 2.4M                              \\
\midrule
With 5\% Dropout \& 79\% Enrollment & & & &\\
\midrule                           
With Washington Park & 384$\rightarrow$268  & 555$\rightarrow$385    & 1.6M   & 2.5M                              \\
Without Washington Park   & 294$\rightarrow$203  & 442$\rightarrow$307 & 1.3M  & 1.9M                              \\
Original Estimate & -                                           & 400$\rightarrow$275                                 & -                                       & 0.5M                                \\
\bottomrule
\end{tabular}
\caption{Estimated 7-year costs of the program for different scenarios and the original program cost
estimate \cite{bed17}. The original estimate did not consider liens.}
\label{res}
\end{table*}

\begin{figure}[h]
\centering
  \includegraphics[trim = 0mm 0mm 0mm 0mm, clip,width=0.475\textwidth]{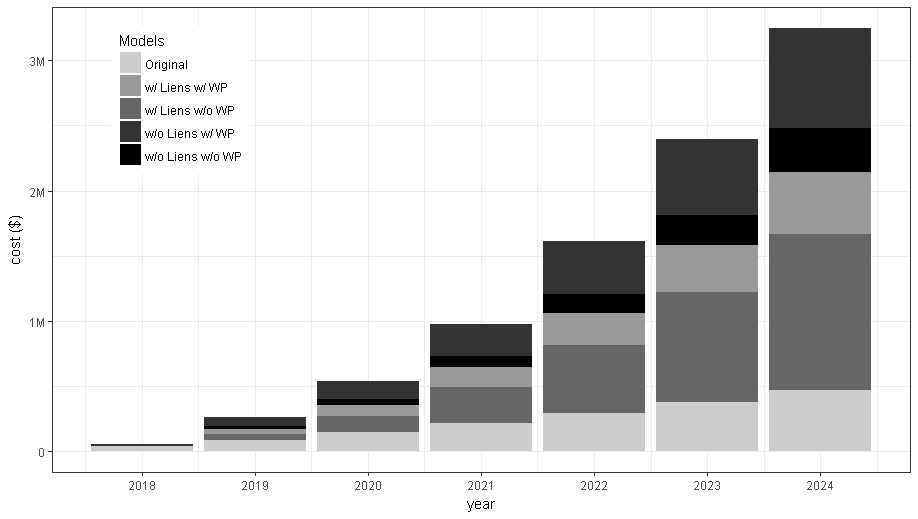} 
\caption{Total program costs for several models, with the 5\% dropout rate and 79\% enrollment which the original model included, during the first 7-year period \cite{bed17}.}
\label{7_year}
\end{figure}

We show that the initial study of the Anti-Displacement Tax Fund significantly underestimated the number of eligible participants and the total program cost, even though we include a large lien rate, which disqualified 40\% of the otherwise eligible homeowners (see Table \ref{res} and Figure \ref{7_year}). While our estimates are limited to 7 years due to our home value appreciation modeling, the increased number of eligible homeowners and larger property value appreciation rates in our evaluation will drastically increase the program cost over 20 years compared to the initial projection. Furthermore, if community members can convince the ADTF to include the Washington Park neighborhood, the number of eligible participants and the total program cost will be much larger than the original estimate (see Table \ref{res} and Figure \ref{7_year}).

These differences in program cost estimates are primarily due to our use of granular data; using individual household-level data for a program aimed at individual homeowners is more appropriate for this type of program evaluation. Identifying which homes were occupied by the homeowner had a significant impact on the number of eligible homes. By hand checking a sample of households for lien data in the affected neighborhoods, we were able to identify a very large lien rate for homeowners in the Westside neighborhoods. Using machine learning techniques on the filtered CEX data and Zillow rent estimates for homes in our study, we could accurately estimate the income for homeowners in the Westside.

Our program estimates also benefited from the use of more advanced statistical techniques that are generally not applied to program evaluation. Our clustering of the tax assessment time series data revealed significant variation within property value appreciation for the Old Fourth Ward neighborhood after the announcement and construction of the nearby Beltline. We could identify that urban renewal in the area considerably weakened the impact of the recession on neighborhood property values, albeit with different effects for homes with different characteristics. We identified these pertinent home characteristics from the clusters, even though there were several classic statistical issues with the data, such as nonlinear relationships, and strong correlations between home features. Proximity to the Beltline, the value of the home prior to construction of the Beltline, and the age of the home were the most important characteristics for increased home value appreciation. We also observed property appreciation rates that were larger than the original program study found and had nonlinear increases over time.

\section{Discussion}

This project demonstrates the value of community-driven data science for academia, city planners and community-based advocates alike. In comparison with the original study, our findings show that the estimated cost of the tax fund could be significantly higher than expected. Our model predicts less homeowners will be able to qualify for the program as the previous study found yet we predicted similar costs for the program during the first 7 years. This is due to larger tax assessment appreciation rates and we expect the program to ultimately cost more than the previous projection. Accordingly, we find the use of machine learning techniques to be valuable tools for quantifying similar anti-gentrification initiatives.  This work highlights that machine learning techniques and data driven science can be valuable for measuring the impacts of urban projects. City planners are not often trained in these fields and where these skills are lacking, cities can benefit from academic partnership.

While this exercise in data science proves useful to planners and local philanthropists, our primary partners are community residents. The results of this project support the community in two ways. First, the Fund becomes much more transparent. Residents can use the visual eligibility tool to determine if they are eligible and make an informed decision to participate. Prior to our work, residents interested in applying for the Fund were required to fill out a paper-based form and deliver it in person or fax it to the Westside Future Fund. We anticipate that many residents may not have access to web-based tools; however, community leaders have demonstrated interest in facilitating use of the online tool for such residents. Not only is the process of qualification more transparent, but the research we conducted to inform our modeling will also inform the community of what the Anti-Displacement Tax Fund entails. Take for example the lien eligibility requirement. Property lien status is key component of the eligibility requirements. While it is just preliminary results, our research and modeling reveals that many of the homes within the five neighborhoods have liens associated with them. This dramatically reduces the number of residents that are eligible to receive help from the fund, which in turn decreases the overall value and scope of the tax fund. With this knowledge the community residents could advocate that part of the tax fund could be used to pay off liens within a specified threshold.

Quantifying the program also allows residents to offer alternative solutions. The community has already voiced two primary concerns for why the anti-displacement tax fund will not fully address displacement and having the fund quantified with reliable data on who qualifies allows for greater legitimacy as they continue to advocate for ways to improve the impact of the fund or alternatives like permanently affordable housing through the community land trust model. The tradeoff between how many residents are actually eligible to participate and the projected length of the program are also factors to consider. It needs to be decided whether longevity for fewer residents is more important, or if having as many residents as possible participate for a shorter time is the best solution. Either way, the data and visual eligibility tool become fodder for community groups to continue the dialogue on how to achieve development without displacement for as many residents as possible.

\subsection{Future Work}

Our work outlines methods to inform and engage community members in the decision making and planning process, but our predictive modeling is limited by time and the availability of data. Following the inauguration of the Anti-Displacement Tax Fund in 2018, much more data should be available about the number of enrolled households, and program cost and scope will be easier to predict. Recalculating the projected costs of the program each year is essential to avoid surprises and keep the program functioning for as long as possible. Additionally, it is vital that future predictions be made public and accessible to community members so that they can advocate for their rights and be more aware of how the tax fund and other anti-displacement measures are shaping their rapidly transforming community.

A Member of the Westside Atlanta Land Trust has been trained to collect additional lien data from the County Clerk's records. If the mapping tool is used to advertise the ATDF and collect more accurate eligibility information from homeowners in the affected neighborhood, these data will be used to identify if there are significant differences in lien rates between neighborhoods, update the eligibility estimates, and recalculate the cost of the program. Data collection can go beyond the scope of program evaluation. This canvassing could also collect information used for longitudinal studies of urban renewal projects and their relationship to displacement, such as knowing where previous residents have moved to and why.

Finally, it would be beneficial to conduct an event study of the Mercedes-Benz stadium construction, as we were only able to include the estimated impact of the Beltline in our analysis and predictions. It is likely that the stadium's construction has already had an impact on home sales and prices in the Westside area. These effects should be incorporated into a future model to more accurately predict property tax increases and the eventual cost of the program.

{\bf Acknowledgments.} We would like to thank Myeong Lee for assistance with developing the online tool. We are also grateful for the continued feedback and support from the Westside Atlanta Land Trust, a program of HELP Org. 

{\bf Research Support.} This work was supported by the NSF IIS \#1659757.




\nocite{*}
\bibliographystyle{abbrv}
\bibliography{references}

\end{document}